\documentclass[aip,jap,reprint,floatfix]{revtex4-1}
\usepackage{graphicx}
\usepackage{footnote}
\usepackage{amsmath}

\begin{document}


\title{Resonant and Time-Resolved Spin Noise Spectroscopy}%

\author{Brennan C. Pursley}%
\email{bpursley@umich.edu}
\affiliation{Applied Physics Program, University of Michigan, Ann Arbor, MI 48109}

\author{X. Song}%
\affiliation{Applied Physics Program, University of Michigan, Ann Arbor, MI 48109}

\author{V. Sih}
\affiliation{Applied Physics Program, University of Michigan, Ann Arbor, MI 48109}
\affiliation{Department of Physics, University of Michigan, Ann Arbor, MI 48109}

\date{\today}

\begin{abstract}
We demonstrate a method to extend the range of pulsed laser spin noise measurements to long spin lifetimes.  We use an analog time-domain detection scheme with a bandwidth limited only by laser pulse duration.  Our model uses statistics and Bloch-Torrey equations to extract the Lande g-factor, Faraday cross-section $\sigma_F$, and spin lifetime $\tau_S$, while accounting for finite detector response. Varying the magnetic field with a fixed probe-probe delay yields $\tau_S$ when it is longer than the laser repetition period.  Varying the probe-probe delay with a fixed field produces a time-domain measurement of the correlation function.
\end{abstract}


\maketitle

Noise is an expression of a system's fundamental behavior.  The fluctuation-dissipation theorem,\cite{Weber1956,Kubo1966} when combined with the Wiener-Khinchin theorem,\cite{Cohen1998} implies that a system's dynamic response can be extracted from correlation measurements.  Spin noise was theoretically postulated in 1946 \cite{Bloch1946a} and then demonstrated on atomic gases, \cite{Aleksandrov1981,Sørensen1998,Crooker2004} nuclear spins,\cite{Sleator1985} and with scanning tunneling microscopy.\cite{Nussinov2003}  More recently, spin noise spectroscopy on semiconductors \cite{Oestreich2005,Crooker2009} has revealed its potential for performing contactless and non-perturbative measurements that can also more directly access the Faraday cross-section.\cite{Giri2012}  Experiments that utilize spin noise as a level of contrast have performed non-destructive spatial resolution of a semiconductor's dopant density \cite{Romer2009} and homogeneous absorption linewidth measurements of quantum dot ensembles.\cite{Yang2014}

Conventional spin noise spectroscopy utilizes a continuous wave laser transmitted through a sample to measure the Faraday rotation produced by stochastic spin fluctuations.  A magnetic field is applied orthogonal to the optical path to induce precession and measure the Lande g-factor.  A power spectrum is produced either by using electronic sampling and Fourier transforming the data set\cite{Muller2010,Zapasskii2013b} or using a spectrum analyzer.\cite{Oestreich2005,Huang2011}  Isolation of the spin noise contribution from a typically much larger background is accomplished by taking the difference of the finite power spectra for two system states.  In these measurements, the bandwidth is limited by the sampling rate.

In order to access high-frequency dynamics, theoretical \cite{Starosielec2008} and experimental \cite{Muller2010b,Hubner2013,Berski2013a} efforts have included the use of ultrafast pulsed lasers.  Ultrafast spin noise spectroscopy allows for a direct time-domain measurement of the spin correlation function, with a bandwidth determined by the laser pulse duration.\cite{Muller2010b,Hubner2013}  However, the temporal range is limited by the optical delay time between probe pulses and, ultimately, the laser repetition period $t_{rep}$, which makes it difficult to accurately determine the spin lifetime $\tau_S$ when $\tau_S \gtrsim t_{rep}$.  In this Letter, we demonstrate a Resonant Spin Noise (RSN) technique that enables accurate measurements of the spin lifetime when $\tau_S \gtrsim t_{rep}$ and clarify how to theoretically model the ultrafast spin noise signal and extract the Faraday cross-section.  We compare our data and model with the model presented in Ref. 21\nocite{Berski2013a} which predicts a Lorentzian peak at zero magnetic field that we do not observe.

We perform ultrafast spin noise measurements using analog electronics. We process our signal with an analog root-mean-square (RMS) circuit and then isolate the spin noise component with lock-in amplification.  We develop a model to explain the analog electronic calculation based on statistics and simplified Bloch-Torrey equations which are derivable from the semi-classical Boltzmann equation.\cite{Bloch1946a,Torrey1956,Qi2003}  By including the response time of our circuitry, we are able to quantify our signal amplitude and extract the Faraday cross-section.\cite{Giri2012}  We show that when the spin lifetime $\tau_S$ is short compared to the laser repetition period $t_{rep}$, measurements as a function of relative probe delay $\Delta t$ accurately capture the spin dynamics.  When $\tau_S \gtrsim t_{rep}$, we show that measurements as a function of applied magnetic field with fixed $\Delta t$ accurately determine $\tau_S$.

A system's dynamics can be obtained from noise through the correlation of two measurements $\theta_i$ and $\theta_j$.  In practice, the correlation function is obtained from measurements of the expectation value of the product of $\theta_i$ and $\theta_j$:
\begin{equation}
\left\langle \theta_i \theta_j \right\rangle = \left\langle \theta_i \right\rangle \left\langle \theta_j \right\rangle + Var\left(\theta \right) Corr\left(\theta_i,\theta_j \right)
\label{eq: expectation-of-product}
\end{equation}
Here, we separate the correlation into the product of its magnitude, $Var\left( \theta \right)$, the variance of measurements $\theta_i$, and a normalized correlation function $Corr\left(\theta_i,\theta_j\right)$.  For a sample with no net spin polarization, $\left\langle \theta_i \right\rangle = \left\langle \theta_j \right\rangle = 0$.

To acquire $\left\langle \theta_i \theta_j \right\rangle$, we can sum $\theta_i$ and $\theta_j$, square the sum, and then perform an average of multiple squared sums.  We accomplish these operations using a series of analog electronic components including a low pass filter and the RMS channel of a telecommunications chip (Analog Devices ADL5511-EVALZ).  Due to the analog processing, the first sum turns out to be a weighted average (WA) of many $\theta_i$, as explained in the Supplementary Material.\footnote{\label{nt: supMat}See Appendices for a schematic of the optical path and equation derivations.}  To isolate the spin contribution to the signal $\theta$ from the background $\xi$, we use lock-in detection, with the result
\begin{align}
RMS \: Signal &= \left\langle \left( \theta+\xi \right)_{WA}^2 \right\rangle^{1/2} - \left\langle \xi_{WA}^2 \right\rangle^{1/2} \nonumber \\ 
&\simeq \left\langle \theta_{WA}^2 \right\rangle^{1/2}
\label{eq: signal-root}
\end{align}
In the second line, we have made the assumption that $\theta \gg \xi$ which, for spin noise measurements of n-GaAs, can be accomplished through the use of sufficient probe power, although too much power will perturb the sample.\cite{Crooker2009}  $\left\langle \theta_{WA}^2 \right\rangle$ is defined as:
\begin{equation}
\left\langle \theta_{WA}^2 \right\rangle \simeq \left(\int_{0}^{\delta} dt \, \frac{e^{-t / t_{RC}}}{\delta} \right)^2 \frac{\sum_{i,j}^{\infty} e^{-(t_i + t_j) / t_{RC}} \left\langle \theta_i \theta_j \right\rangle }{\sum_{i,j}^{\infty} e^{-(t_i + t_j) / t_{RC}}}
\label{eq: theta-wa-squared}
\end{equation}
The leading integral averages the voltage decay between the most recent probe pulse and the next probe pulse where $t_{RC}$ is the electronic response time.  The discrete sums provide a weighted average over all previous pulse residues up to, and including, the most recent pulse.

Our model\footnotemark[\value{footnote}] for carrier spin dependent behavior in n-GaAs is
\begin{equation}
\left\langle \theta_i \theta_j \right\rangle = Var\left( \theta \right) e^{-\vert t_i - t_j \vert / \tau_S} cos\left(\Omega \vert t_i - t_j \vert \right)
\label{eq: probe-probe-expectation}
\end{equation}
with $Var\left( \theta \right) = \sigma_F n d / 16 A $ where $n$ is the sample carrier density, $d$ is the sample thickness, and $A$ is the probe cross-section at the sample.  $\sigma_F$ is the Faraday cross-section where $\theta = \sigma_F n_S d$ with $n_S = n_{\uparrow} - n_{\downarrow} = Pn$ being the sample's spin polarized carrier density with $P$ the level of polarization.\cite{Giri2012}  $\Omega = \mu_B g B / \hbar$ is the Larmor precession frequency where $\mu_B$ is the Bohr magneton, $g$ is the Lande g-factor, $\hbar$ is the reduced Planck's constant, and $B$ is the applied magnetic field strength.

We interpret the Bloch-Torrey equations as deriving the ensemble correlation function, a normalized description of the dynamics.  Our function $Var\left(\theta\right)$ contains all information about the amplitude and is calculated by taking an RMS over all initial ensemble states.  As the Bloch-Torrey equations are semi-classical and vectorial, we represent the two-state system for electrons with a binomial distribution.  We also average over all possible orientations on the unit sphere with the assumption that there is no preferential axis of quantization.

Our experiment utilizes a 76 MHz repetition rate ($t_{rep} \simeq 13.6$ ns) Ti:Sapphire laser with 3 ps pulse duration.  We pass the output through a beam splitter (BS) to create fixed and mechanically variable length (VL) paths with separate pulse trains.  We recombine the paths at a second beam splitter prior to entering a fiber optic cable (FOC).  After exiting the FOC and passing through a linear polarizer, the two pulse trains will be identical in polarization, intensity, and spatial profile prior to interaction with the sample where they undergo Faraday rotation proportional to the stochastic spin polarization.  Each pulse train has an average power of 10 mW and is focused to an area of 500 $\mu$m$^2$ at the sample.

We use an electro-optic modulator (EOM) to turn on and off our ability to detect the Faraday rotation of pulses transmitted through the sample, as described in Ref. 21\nocite{Berski2013a}.  This is achieved by either allowing unaltered linear polarization to proceed (EOM is at zero retardance) or changing the linear polarization to circular polarization (EOM is at quarter-wave retardance) with a 157 Hz square-wave switching frequency.  After the EOM, the light passes through a Wollaston prism (WP) which splits the pulses into orthogonal linear components.  When the EOM is at zero retardance, a Faraday rotation will be detected as an imbalance at the photodiode bridge (PDB).  At quarter-wave retardance, no imbalance will be detected.  After the PDB, the photogenerated voltage is amplified, and then processed on the RMS circuit of the ADL5511-EVALZ.  The limited speed of our PDB, combined with parasitic capacitances in our wiring, provide an effective low-pass filter to generate a weighted average with $t_{RC} \sim29$ ns.  Final detection of our modulated signal occurs at a lock-in amplifier (LIA).

\begin{figure*}
\includegraphics[scale=1]{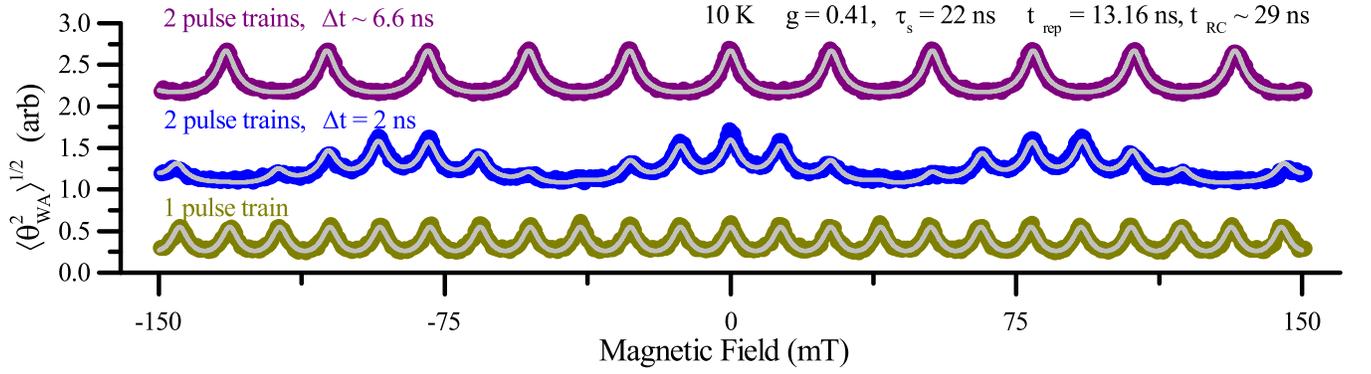}
\caption{Resonant Spin Noise (RSN) data taken at 10 K with fits shown as solid gray lines.  Single pulse train data is fit using Eqs. \ref{eq: signal-root}, \ref{eq: theta-wa-squared}, and \ref{eq: probe-probe-expectation} while two pulse train data is fit by Eqs. \ref{eq: probe-probe-expectation} and \ref{eq: variable-spacing}.  Data are scaled and offset for clarity.}
\label{fig: Fig1}
\end{figure*}

\begin{figure*}
\includegraphics[scale=1]{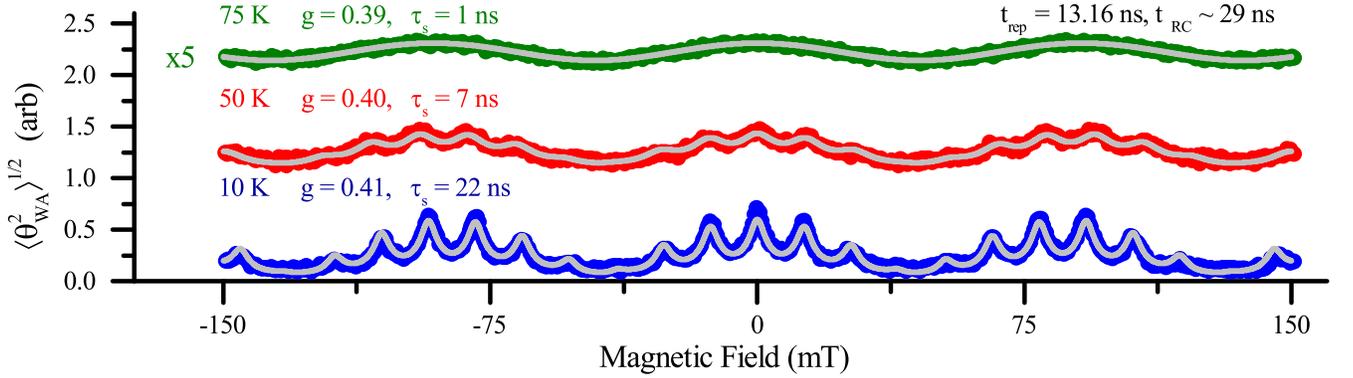}
\caption{Resonant Spin Noise (RSN) data taken at various temperatures with fits using Eqs. \ref{eq: probe-probe-expectation} and \ref{eq: variable-spacing} shown as solid gray lines.  $\vert \Delta t \vert = 2$ ns for all data.  Data are scaled and offset for clarity.}
\label{fig: Fig2}
\end{figure*} 

We tested our measurement scheme by performing spin noise measurements on a 500 $\mu$m thick chip of bulk Si-doped n-type $\left\langle001\right\rangle$ GaAs.  The following are the manufacturer specifications: carrier concentration of $(4.3-6.2)\times10^{16}$ cm$^{-3}$, mobility of (3450-3880) cm$^2$/V$\cdot$s, and resistivity of $(2.8-3.9)\times10^{-2}$ Ohm$\cdot$cm.  The electron carrier concentration of $\sim$ $5\times10^{16}$ cm$^{-3}$ is near the metal-to-insulator transition of GaAs where long spin lifetimes are expected.\cite{Dzhioev2002}
  
Figures \ref{fig: Fig1} and \ref{fig: Fig2} show Resonant Spin Noise (RSN), in which the delay between two pulse trains remains fixed while the magnetic field is swept.  The resultant signal is reminiscent of the pump-probe method of Resonant Spin Amplification (RSA) \cite{Kikkawa1998} but arises from a different measurement principle.  RSA is the result of multiple induced spin populations with integer phase relationships undergoing precession that are all probed at the same point in time.  RSN is the precession of a single stochastic spin population, probed at a series of times that undergo subsequent processing to extract correlations.  Due to its minimally perturbative capabilities, RSN, and spin noise more generally, should provide simpler analysis of $\tau_S$ compared to values extracted with RSA if the optically-pumped spin polarization is affected by subsequent pump pulses, as observed in Ref. 27\nocite{Kuhlen2014}.

We compare data collected at 10 K for a single pulse train and two pulse trains with different delays.  Using a relative delay ($\Delta t \sim 6.6$ ns = $t_{rep}/2$) that is half of the laser repetition period leads to a signal (Fig. \ref{fig: Fig1}, top) that is similar to a single pulse train (Fig. \ref{fig: Fig1}, bottom).  This is due to the equal time spacing of the probe pulses.  Equations \ref{eq: signal-root}, \ref{eq: theta-wa-squared}, and \ref{eq: probe-probe-expectation} are used to model the behavior of the single pulse train data, shown as a solid gray line, with $\delta = t_{rep}$ and $t_i - t_j = (i-j)t_{rep}$. Letting $t_{rep} \rightarrow t_{rep}/2$  allows for fitting of the two pulse train case. 

\begin{figure}
\includegraphics[scale=1]{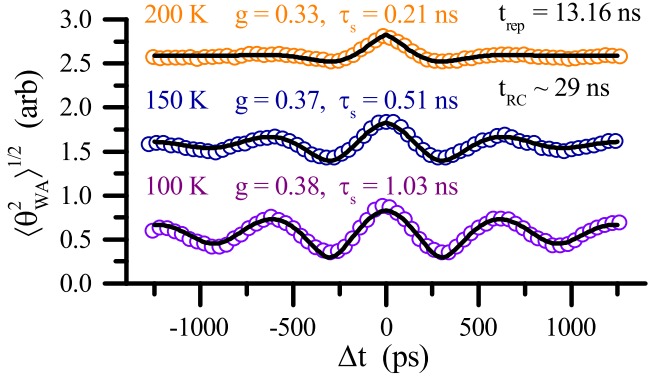}
\caption{Time-Resolved Spin Noise (TRSN) data taken at various temperatures with fits using Eqs. \ref{eq: probe-probe-expectation} and \ref{eq: variable-spacing} shown as solid black lines.  The applied magnetic field strength was 300 mT for all data shown.  Data are scaled and offset for clarity.}
\label{fig: Fig3}
\end{figure}

For probe pulses that are not equally spaced, we must modify Eq. \ref{eq: theta-wa-squared}.  Equation \ref{eq: variable-spacing} is a weighted average of the two possible timing sequences that stem from the first-in, first-out behavior of our electronic processing.
\begin{subequations}
\label{eq: variable-spacing}
\begin{gather}
\left\langle \theta_{WA}^2 \right\rangle = \frac{\beta}{t_{rep}}\sqrt{f_{\alpha,\beta}} + \frac{\alpha}{t_{rep}}\sqrt{f_{\beta,\alpha}} \label{eq: variable-spacing-a} \\
f_{\alpha,\beta} = \left( \int_{0}^{\beta} dt \, \frac{e^{-t / t_{RC}}}{\beta} \right)^2 \frac{\sum_{i,j}^{\infty} e^{-(t_i^{\alpha} + t_j^{\alpha}) / t_{RC}} \left\langle \theta_i^{\alpha} \theta_j^{\alpha} \right\rangle }{\sum_{i,j}^{\infty} e^{-(t_i^{\alpha} + t_j^{\alpha}) / t_{RC}}} \label{eq: variable-spacing-b}
\end{gather}
\end{subequations}
with $t_i^{\alpha} \in \left\lbrace 0,\alpha,t_{rep}, t_{rep} + \alpha, ... \right\rbrace$, $\alpha=\vert \Delta t \vert$, and $\beta = t_{rep}-\vert \Delta t \vert$ where $\vert \Delta t \vert \leq t_{rep}$ is the relative spacing between the two pulse trains controlled by the variable length optical delay line.  Equations \ref{eq: probe-probe-expectation} and \ref{eq: variable-spacing} are used to fit all two pulse train data, including Time-Resolved Spin Noise (TRSN) shown in Fig. \ref{fig: Fig3}.  We calibrated our system using $\vert \Delta t \vert = 2$ ns RSN data, combined with $g$ and $\tau_S$ values extracted from 10 K RSA measurements, yielding $t_{RC} \sim 29$ ns.  It should be noted that this leaves only three parameters for the rest of the fits: $g$, $\tau_S$, and $\sigma_F$.

Figure \ref{fig: Fig2} shows the temperature dependent evolution of RSN data with $\vert \Delta t \vert = 2$ ns up to 75 K.  As temperature increases, $\tau_S$ decreases as well as the contribution of terms with larger temporal spacing in Eq. \ref{eq: variable-spacing}.  By 75 K, $\tau_S$ is small enough that the error in fitting RSN is appreciable.  However, the 75 K data is a useful qualitative test of our model.  We do not observe a Lorentzian peak at zero magnetic field as expected from the antiderivative of Eq. 1 in Ref. 21\nocite{Berski2013a}.\footnotemark[\value{footnote}]

To obtain a more accurate measurement of the spin lifetime for shorter $\tau_S$, we can instead perform TRSN where we scan the time delay $\Delta t$ between the pulses at a fixed magnetic field.  In Time-Resolved Faraday Rotation (TRFR),\cite{Baumberg1994} a pump pulse orients a single population of spins and a delayed probe measures their evolved orientation.  In TRSN, a single population of stochastically aligned spins is probed by an initial pulse and then compared to a delayed pulse through Eq. \ref{eq: probe-probe-expectation}, capturing the system’s evolution.

Figure \ref{fig: Fig3} shows TRSN at 100, 150 and 200 K with an applied magnetic field of 300 mT.  Equations \ref{eq: probe-probe-expectation} and \ref{eq: variable-spacing} are used for fitting and shown as solid black lines.  If $\vert \Delta t \vert, \tau_S \ll t_{RC}, t_{rep}$, we can simplify Eqs. \ref{eq: probe-probe-expectation} and \ref{eq: variable-spacing} to obtain Eq. \ref{eq: theta-WA-sq-simple}.

\begin{equation}
\left\langle \theta_{WA}^2 \right\rangle^{1/2} \sim \sqrt{1 + e^{-\vert \Delta t \vert / \tau_S} cos\left(\Omega \vert \Delta t \vert \right)}
\label{eq: theta-WA-sq-simple}
\end{equation}

\begin{figure}
\includegraphics[scale=1]{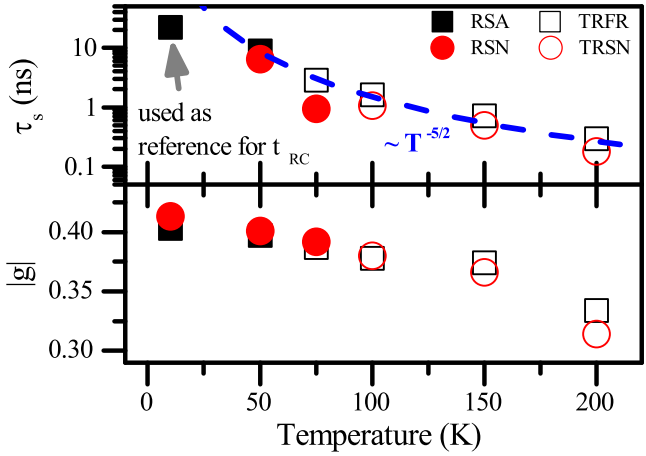}
\caption{Temperature dependence of fit values extracted from Resonant Spin Noise (RSN) and Time Resolved Spin Noise (TRSN) data shown in Figs. \ref{fig: Fig3} and \ref{fig: Fig4} compared with fit values extracted from the conventional techniques of Resonant Spin Amplification (RSA) and Time Resolved Faraday Rotation (TRFR).  The blue dashed line is a power law fit, $\sim T^{-5/2}$, for the D'yakonov-Perel dephasing mechanism.}
\label{fig: Fig4}
\end{figure}

Figure \ref{fig: Fig4} shows the temperature dependence of the fit values $g$ and $\tau_S$ determined by RSA and TRFR and our spin noise methods of RSN and TRSN.  The agreement is quite good, with the exception of the RSN value for $\tau_S$ at 75 K, and is consistent with previous reports.\cite{Kikkawa1998,Oestreich1995,Hubner2009}  We have included a blue dashed line in Fig. \ref{fig: Fig4} showing the expected D'yakanov-Perel temperature dependence for $\tau_S$ which, for $T\geq 50$ K, follows the power law $\tau_S \sim T^{-5/2}$.

\begin{figure}
\includegraphics[scale=1]{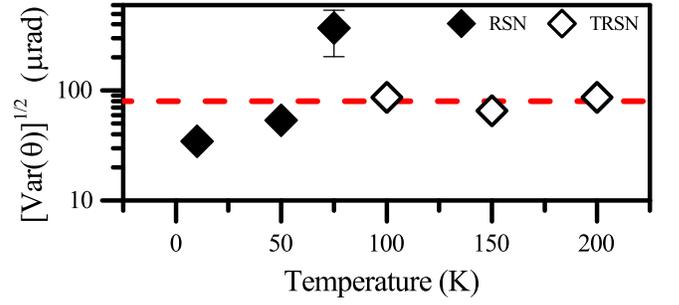}
\caption{Temperature dependence of $Var\left(\theta\right)^{1/2}$ extracted from RSN and TRSN data.  The red dashed line is the average of $Var\left(\theta\right)^{1/2}$ for 100 K through 200 K used to extract a value for $\sigma_F$.}
\label{fig: Fig5}
\end{figure}

Figure \ref{fig: Fig5} shows the temperature dependence of $\left[Var\left(\theta\right)\right]^{1/2}$ extracted from RSN and TRSN data.  Since our sample is near the metal-to-insulator transition and the energy level of Si donors in GaAs is $\sim 6$ meV below the conduction band,\cite{Yu2010} the majority of Si donors should be ionized above 100 K.  We observe nearly constant behavior for $\left[Var\left(\theta\right)\right]^{1/2}$ above 100 K and attribute this to a constant carrier density, which results in a constant value for $\sigma_F$.  We use the average value of $\left[Var\left(\theta\right)\right]^{1/2}$ above 100 K, along with $n = 5\times10^{16}$ cm$^{-3}$, $d = 500$ $\mu$m, and $A$ = 500 $\mu$m$^2$ to extract $\sigma_F$ yielding the value $\sigma_F \simeq 1 \times 10^{-14} $  rad$\cdot$cm$^2$ in agreement with Ref. 12\nocite{Giri2012}.  Moreover, our measurement does not require any assumptions regarding optical pumping efficiency.

Spin noise techniques can also offer advantages when attempting to study material systems with inefficient optical orientation.  The signal strength of conventional pump-probe techniques relies upon spin dependent optical selection rules, as well as sample dimensions and probe spot size.  We can determine the approximate threshold where spin noise provides a stronger signal through the ratio of $\left[Var\left(\theta\right)\right]^{1/2}$ to the maximum pump-probe Faraday rotation $\theta_{PP} = \sigma_F P n d$:
\begin{equation}
\frac{\left[Var\left(\theta\right)\right]^{1/2}}{\theta_{PP}} = \frac{1}{4 P \sqrt{n d A}}
\end{equation}
The dimensionless maximum polarization value P is determined by selection rule efficiency and experiment conditions.  If $P \leq 1/4\sqrt{n d A}$, then spin noise should be a more advantageous experiment.  We can use our result to evaluate the interesting case of probing a single spin where $n d A = 1$.  Spin noise should allow the observation of dynamics, without the need to filter out pump scatter, with signal that is on the order of maximum achievable pump-probe Faraday rotation.

In conclusion, we have demonstrated Resonant Spin Noise (RSN) and Time Resolved Spin Noise (TRSN).  Both methods utilize the same optical path and electronics which, in principle, offers a combined bandwidth beyond 1 THz with commercially available femtosecond pulsed lasers.  We model our analog processing and explored spin dynamics using statistics and simplified Bloch-Torrey equations.  We interpret the Bloch-Torrey equations as deriving the correlation function while the amplitude of the signal is the root-mean-square of all possible system states.  From fitting, we extract values for the dephasing time, Lande g-factor, and Faraday cross-section that agree with conventional measurements and available literature.

This work was supported by the Defense Threat Reduction Agency Basic Research Award No. HDTRA1-13-1-0013, the National Science Foundation Materials Research Science and Engineering Center program DMR-1120923, the Office of Naval Research, and the Air Force Office of Scientific Research.  B.P. was supported in part by the Graduate Student Research Fellowship under Grant No. DGE-1256260.

\bibliographystyle{aipnum4-1}
\bibliography{ReferencesV7}

\begin{thebibliography}{31}%
\makeatletter
\providecommand \@ifxundefined [1]{%
 \@ifx{#1\undefined}
}%
\providecommand \@ifnum [1]{%
 \ifnum #1\expandafter \@firstoftwo
 \else \expandafter \@secondoftwo
 \fi
}%
\providecommand \@ifx [1]{%
 \ifx #1\expandafter \@firstoftwo
 \else \expandafter \@secondoftwo
 \fi
}%
\providecommand \natexlab [1]{#1}%
\providecommand \enquote  [1]{``#1''}%
\providecommand \bibnamefont  [1]{#1}%
\providecommand \bibfnamefont [1]{#1}%
\providecommand \citenamefont [1]{#1}%
\providecommand \href@noop [0]{\@secondoftwo}%
\providecommand \href [0]{\begingroup \@sanitize@url \@href}%
\providecommand \@href[1]{\@@startlink{#1}\@@href}%
\providecommand \@@href[1]{\endgroup#1\@@endlink}%
\providecommand \@sanitize@url [0]{\catcode `\\12\catcode `\$12\catcode
  `\&12\catcode `\#12\catcode `\^12\catcode `\_12\catcode `\%12\relax}%
\providecommand \@@startlink[1]{}%
\providecommand \@@endlink[0]{}%
\providecommand \url  [0]{\begingroup\@sanitize@url \@url }%
\providecommand \@url [1]{\endgroup\@href {#1}{\urlprefix }}%
\providecommand \urlprefix  [0]{URL }%
\providecommand \Eprint [0]{\href }%
\providecommand \doibase [0]{http://dx.doi.org/}%
\providecommand \selectlanguage [0]{\@gobble}%
\providecommand \bibinfo  [0]{\@secondoftwo}%
\providecommand \bibfield  [0]{\@secondoftwo}%
\providecommand \translation [1]{[#1]}%
\providecommand \BibitemOpen [0]{}%
\providecommand \bibitemStop [0]{}%
\providecommand \bibitemNoStop [0]{.\EOS\space}%
\providecommand \EOS [0]{\spacefactor3000\relax}%
\providecommand \BibitemShut  [1]{\csname bibitem#1\endcsname}%
\let\auto@bib@innerbib\@empty
\bibitem [{\citenamefont {Weber}(1956)}]{Weber1956}%
  \BibitemOpen
  \bibfield  {author} {\bibinfo {author} {\bibfnamefont {J.}~\bibnamefont
  {Weber}},\ }\href {\doibase 10.1103/PhysRev.101.1620} {\bibfield  {journal}
  {\bibinfo  {journal} {Phys. Rev.}\ }\textbf {\bibinfo {volume} {101}},\
  \bibinfo {pages} {1620} (\bibinfo {year} {1956})}\BibitemShut {NoStop}%
\bibitem [{\citenamefont {Kubo}(1966)}]{Kubo1966}%
  \BibitemOpen
  \bibfield  {author} {\bibinfo {author} {\bibfnamefont {R.}~\bibnamefont
  {Kubo}},\ }\href {http://iopscience.iop.org/0034-4885/29/1/306} {\bibfield
  {journal} {\bibinfo  {journal} {Rep. Prog. Phys.}\ }\textbf {\bibinfo
  {volume} {29}},\ \bibinfo {pages} {255} (\bibinfo {year} {1966})}\BibitemShut
  {NoStop}%
\bibitem [{\citenamefont {Cohen}(1998)}]{Cohen1998}%
  \BibitemOpen
  \bibfield  {author} {\bibinfo {author} {\bibfnamefont {L.}~\bibnamefont
  {Cohen}},\ }\href {\doibase 10.1109/97.728471} {\bibfield  {journal}
  {\bibinfo  {journal} {IEEE Signal Processing Lett.}\ }\textbf {\bibinfo
  {volume} {5}},\ \bibinfo {pages} {292} (\bibinfo {year} {1998})}\BibitemShut
  {NoStop}%
\bibitem [{\citenamefont {Bloch}(1946)}]{Bloch1946a}%
  \BibitemOpen
  \bibfield  {author} {\bibinfo {author} {\bibfnamefont {F.}~\bibnamefont
  {Bloch}},\ }\href {\doibase 10.1103/PhysRev.70.460} {\bibfield  {journal}
  {\bibinfo  {journal} {Phys. Rev.}\ }\textbf {\bibinfo {volume} {70}},\
  \bibinfo {pages} {460} (\bibinfo {year} {1946})}\BibitemShut {NoStop}%
\bibitem [{\citenamefont {Aleksandrov}\ and\ \citenamefont
  {Zapasskil}(1981)}]{Aleksandrov1981}%
  \BibitemOpen
  \bibfield  {author} {\bibinfo {author} {\bibfnamefont {E.~B.}\ \bibnamefont
  {Aleksandrov}}\ and\ \bibinfo {author} {\bibfnamefont {V.~S.}\ \bibnamefont
  {Zapasskil}},\ }\href@noop {} {\bibfield  {journal} {\bibinfo  {journal}
  {Sov. Phys. JETP}\ }\textbf {\bibinfo {volume} {54}},\ \bibinfo {pages} {64}
  (\bibinfo {year} {1981})}\BibitemShut {NoStop}%
\bibitem [{\citenamefont {S\o{}rensen}, \citenamefont {Hald},\ and\
  \citenamefont {Polzik}(1998)}]{Sørensen1998}%
  \BibitemOpen
  \bibfield  {author} {\bibinfo {author} {\bibfnamefont {J.~L.}\ \bibnamefont
  {S\o{}rensen}}, \bibinfo {author} {\bibfnamefont {J.}~\bibnamefont {Hald}}, \
  and\ \bibinfo {author} {\bibfnamefont {E.~S.}\ \bibnamefont {Polzik}},\
  }\href {\doibase 10.1103/PhysRevLett.80.3487} {\bibfield  {journal} {\bibinfo
   {journal} {Phys. Rev. Lett.}\ }\textbf {\bibinfo {volume} {80}},\ \bibinfo
  {pages} {3487} (\bibinfo {year} {1998})}\BibitemShut {NoStop}%
\bibitem [{\citenamefont {Crooker}\ \emph {et~al.}(2004)\citenamefont
  {Crooker}, \citenamefont {Rickel}, \citenamefont {Balatsky},\ and\
  \citenamefont {Smith}}]{Crooker2004}%
  \BibitemOpen
  \bibfield  {author} {\bibinfo {author} {\bibfnamefont {S.~A.}\ \bibnamefont
  {Crooker}}, \bibinfo {author} {\bibfnamefont {D.~G.}\ \bibnamefont {Rickel}},
  \bibinfo {author} {\bibfnamefont {a.~V.}\ \bibnamefont {Balatsky}}, \ and\
  \bibinfo {author} {\bibfnamefont {D.~L.}\ \bibnamefont {Smith}},\ }\href
  {\doibase 10.1038/nature02804} {\bibfield  {journal} {\bibinfo  {journal}
  {Nature}\ }\textbf {\bibinfo {volume} {431}},\ \bibinfo {pages} {49}
  (\bibinfo {year} {2004})}\BibitemShut {NoStop}%
\bibitem [{\citenamefont {Sleator}\ \emph {et~al.}(1985)\citenamefont
  {Sleator}, \citenamefont {Hahn}, \citenamefont {Hilbert},\ and\ \citenamefont
  {Clarke}}]{Sleator1985}%
  \BibitemOpen
  \bibfield  {author} {\bibinfo {author} {\bibfnamefont {T.}~\bibnamefont
  {Sleator}}, \bibinfo {author} {\bibfnamefont {E.~L.}\ \bibnamefont {Hahn}},
  \bibinfo {author} {\bibfnamefont {C.}~\bibnamefont {Hilbert}}, \ and\
  \bibinfo {author} {\bibfnamefont {J.}~\bibnamefont {Clarke}},\ }\href
  {\doibase 10.1103/PhysRevLett.55.1742} {\bibfield  {journal} {\bibinfo
  {journal} {Phys. Rev. Lett.}\ }\textbf {\bibinfo {volume} {55}},\ \bibinfo
  {pages} {1742} (\bibinfo {year} {1985})}\BibitemShut {NoStop}%
\bibitem [{\citenamefont {Nussinov}, \citenamefont {Crommie},\ and\
  \citenamefont {Balatsky}(2003)}]{Nussinov2003}%
  \BibitemOpen
  \bibfield  {author} {\bibinfo {author} {\bibfnamefont {Z.}~\bibnamefont
  {Nussinov}}, \bibinfo {author} {\bibfnamefont {M.~F.}\ \bibnamefont
  {Crommie}}, \ and\ \bibinfo {author} {\bibfnamefont {A.~V.}\ \bibnamefont
  {Balatsky}},\ }\href {\doibase 10.1103/PhysRevB.68.085402} {\bibfield
  {journal} {\bibinfo  {journal} {Phys. Rev. B}\ }\textbf {\bibinfo {volume}
  {68}},\ \bibinfo {pages} {085402} (\bibinfo {year} {2003})}\BibitemShut
  {NoStop}%
\bibitem [{\citenamefont {Oestreich}\ \emph {et~al.}(2005)\citenamefont
  {Oestreich}, \citenamefont {R\"{o}mer}, \citenamefont {Haug},\ and\
  \citenamefont {H\"{a}gele}}]{Oestreich2005}%
  \BibitemOpen
  \bibfield  {author} {\bibinfo {author} {\bibfnamefont {M.}~\bibnamefont
  {Oestreich}}, \bibinfo {author} {\bibfnamefont {M.}~\bibnamefont
  {R\"{o}mer}}, \bibinfo {author} {\bibfnamefont {R.~J.}\ \bibnamefont {Haug}},
  \ and\ \bibinfo {author} {\bibfnamefont {D.}~\bibnamefont {H\"{a}gele}},\
  }\href {\doibase 10.1103/PhysRevLett.95.216603} {\bibfield  {journal}
  {\bibinfo  {journal} {Phys. Rev. Lett.}\ }\textbf {\bibinfo {volume} {95}},\
  \bibinfo {pages} {216603} (\bibinfo {year} {2005})}\BibitemShut {NoStop}%
\bibitem [{\citenamefont {Crooker}, \citenamefont {Cheng},\ and\ \citenamefont
  {Smith}(2009)}]{Crooker2009}%
  \BibitemOpen
  \bibfield  {author} {\bibinfo {author} {\bibfnamefont {S.~A.}\ \bibnamefont
  {Crooker}}, \bibinfo {author} {\bibfnamefont {L.}~\bibnamefont {Cheng}}, \
  and\ \bibinfo {author} {\bibfnamefont {D.~L.}\ \bibnamefont {Smith}},\ }\href
  {\doibase 10.1103/PhysRevB.79.035208} {\bibfield  {journal} {\bibinfo
  {journal} {Phys. Rev. B}\ }\textbf {\bibinfo {volume} {79}},\ \bibinfo
  {pages} {035208} (\bibinfo {year} {2009})}\BibitemShut {NoStop}%
\bibitem [{\citenamefont {Giri}\ \emph {et~al.}(2012)\citenamefont {Giri},
  \citenamefont {Cronenberger}, \citenamefont {Vladimirova}, \citenamefont
  {Scalbert}, \citenamefont {Kavokin}, \citenamefont {Glazov}, \citenamefont
  {Nawrocki}, \citenamefont {Lema\^{\i}tre},\ and\ \citenamefont
  {Bloch}}]{Giri2012}%
  \BibitemOpen
  \bibfield  {author} {\bibinfo {author} {\bibfnamefont {R.}~\bibnamefont
  {Giri}}, \bibinfo {author} {\bibfnamefont {S.}~\bibnamefont {Cronenberger}},
  \bibinfo {author} {\bibfnamefont {M.}~\bibnamefont {Vladimirova}}, \bibinfo
  {author} {\bibfnamefont {D.}~\bibnamefont {Scalbert}}, \bibinfo {author}
  {\bibfnamefont {K.~V.}\ \bibnamefont {Kavokin}}, \bibinfo {author}
  {\bibfnamefont {M.~M.}\ \bibnamefont {Glazov}}, \bibinfo {author}
  {\bibfnamefont {M.}~\bibnamefont {Nawrocki}}, \bibinfo {author}
  {\bibfnamefont {A.}~\bibnamefont {Lema\^{\i}tre}}, \ and\ \bibinfo {author}
  {\bibfnamefont {J.}~\bibnamefont {Bloch}},\ }\href {\doibase
  10.1103/PhysRevB.85.195313} {\bibfield  {journal} {\bibinfo  {journal} {Phys.
  Rev. B}\ }\textbf {\bibinfo {volume} {85}},\ \bibinfo {pages} {195313}
  (\bibinfo {year} {2012})}\BibitemShut {NoStop}%
\bibitem [{\citenamefont {R\"{o}mer}, \citenamefont {H\"{u}bner},\ and\
  \citenamefont {Oestreich}(2009)}]{Romer2009}%
  \BibitemOpen
  \bibfield  {author} {\bibinfo {author} {\bibfnamefont {M.}~\bibnamefont
  {R\"{o}mer}}, \bibinfo {author} {\bibfnamefont {J.}~\bibnamefont
  {H\"{u}bner}}, \ and\ \bibinfo {author} {\bibfnamefont {M.}~\bibnamefont
  {Oestreich}},\ }\href {\doibase 10.1063/1.3098074} {\bibfield  {journal}
  {\bibinfo  {journal} {Appl. Phys. Lett.}\ }\textbf {\bibinfo {volume} {94}},\
  \bibinfo {pages} {112105} (\bibinfo {year} {2009})}\BibitemShut {NoStop}%
\bibitem [{\citenamefont {Yang}\ \emph {et~al.}(2014)\citenamefont {Yang},
  \citenamefont {Glasenapp}, \citenamefont {Greilich}, \citenamefont {Reuter},
  \citenamefont {Wieck}, \citenamefont {Yakovlev}, \citenamefont {Bayer},\ and\
  \citenamefont {Crooker}}]{Yang2014}%
  \BibitemOpen
  \bibfield  {author} {\bibinfo {author} {\bibfnamefont {L.}~\bibnamefont
  {Yang}}, \bibinfo {author} {\bibfnamefont {P.}~\bibnamefont {Glasenapp}},
  \bibinfo {author} {\bibfnamefont {A.}~\bibnamefont {Greilich}}, \bibinfo
  {author} {\bibfnamefont {D.}~\bibnamefont {Reuter}}, \bibinfo {author}
  {\bibfnamefont {A.~D.}\ \bibnamefont {Wieck}}, \bibinfo {author}
  {\bibfnamefont {D.~R.}\ \bibnamefont {Yakovlev}}, \bibinfo {author}
  {\bibfnamefont {M.}~\bibnamefont {Bayer}}, \ and\ \bibinfo {author}
  {\bibfnamefont {S.~A.}\ \bibnamefont {Crooker}},\ }\href {\doibase
  10.1038/ncomms5949} {\bibfield  {journal} {\bibinfo  {journal} {Nat.
  Commun.}\ }\textbf {\bibinfo {volume} {5}},\ \bibinfo {pages} {4949}
  (\bibinfo {year} {2014})}\BibitemShut {NoStop}%
\bibitem [{\citenamefont {M\"{u}ller}\ \emph
  {et~al.}(2010{\natexlab{a}})\citenamefont {M\"{u}ller}, \citenamefont
  {Oestreich}, \citenamefont {R\"{o}mer},\ and\ \citenamefont
  {H\"{u}bner}}]{Muller2010}%
  \BibitemOpen
  \bibfield  {author} {\bibinfo {author} {\bibfnamefont {G.~M.}\ \bibnamefont
  {M\"{u}ller}}, \bibinfo {author} {\bibfnamefont {M.}~\bibnamefont
  {Oestreich}}, \bibinfo {author} {\bibfnamefont {M.}~\bibnamefont
  {R\"{o}mer}}, \ and\ \bibinfo {author} {\bibfnamefont {J.}~\bibnamefont
  {H\"{u}bner}},\ }\href {\doibase 10.1016/j.physe.2010.08.010} {\bibfield
  {journal} {\bibinfo  {journal} {Physica E}\ }\textbf {\bibinfo {volume}
  {43}},\ \bibinfo {pages} {569} (\bibinfo {year}
  {2010}{\natexlab{a}})}\BibitemShut {NoStop}%
\bibitem [{\citenamefont {Zapasskii}(2013)}]{Zapasskii2013b}%
  \BibitemOpen
  \bibfield  {author} {\bibinfo {author} {\bibfnamefont {V.~S.}\ \bibnamefont
  {Zapasskii}},\ }\href {\doibase 10.1364/AOP.5.000131} {\bibfield  {journal}
  {\bibinfo  {journal} {Adv. Opt. Phot.}\ }\textbf {\bibinfo {volume} {5}},\
  \bibinfo {pages} {131} (\bibinfo {year} {2013})}\BibitemShut {NoStop}%
\bibitem [{\citenamefont {Huang}\ and\ \citenamefont
  {Steel}(2011)}]{Huang2011}%
  \BibitemOpen
  \bibfield  {author} {\bibinfo {author} {\bibfnamefont {Q.}~\bibnamefont
  {Huang}}\ and\ \bibinfo {author} {\bibfnamefont {D.~S.}\ \bibnamefont
  {Steel}},\ }\href {\doibase 10.1103/PhysRevB.83.155204} {\bibfield  {journal}
  {\bibinfo  {journal} {Physical Review B}\ }\textbf {\bibinfo {volume} {83}},\
  \bibinfo {pages} {155204} (\bibinfo {year} {2011})}\BibitemShut {NoStop}%
\bibitem [{\citenamefont {Starosielec}\ and\ \citenamefont
  {Hägele}(2008)}]{Starosielec2008}%
  \BibitemOpen
  \bibfield  {author} {\bibinfo {author} {\bibfnamefont {S.}~\bibnamefont
  {Starosielec}}\ and\ \bibinfo {author} {\bibfnamefont {D.}~\bibnamefont
  {Hägele}},\ }\href {\doibase 10.1063/1.2969041} {\bibfield  {journal}
  {\bibinfo  {journal} {Appl. Phys. Lett.}\ }\textbf {\bibinfo {volume} {93}},\
  \bibinfo {pages} {051116} (\bibinfo {year} {2008})}\BibitemShut {NoStop}%
\bibitem [{\citenamefont {M\"{u}ller}\ \emph
  {et~al.}(2010{\natexlab{b}})\citenamefont {M\"{u}ller}, \citenamefont
  {R\"{o}mer}, \citenamefont {H\"{u}bner},\ and\ \citenamefont
  {Oestreich}}]{Muller2010b}%
  \BibitemOpen
  \bibfield  {author} {\bibinfo {author} {\bibfnamefont {G.~M.}\ \bibnamefont
  {M\"{u}ller}}, \bibinfo {author} {\bibfnamefont {M.}~\bibnamefont
  {R\"{o}mer}}, \bibinfo {author} {\bibfnamefont {J.}~\bibnamefont
  {H\"{u}bner}}, \ and\ \bibinfo {author} {\bibfnamefont {M.}~\bibnamefont
  {Oestreich}},\ }\href {\doibase 10.1103/PhysRevB.81.121202} {\bibfield
  {journal} {\bibinfo  {journal} {Phys. Rev. B}\ }\textbf {\bibinfo {volume}
  {81}},\ \bibinfo {pages} {121202} (\bibinfo {year}
  {2010}{\natexlab{b}})}\BibitemShut {NoStop}%
\bibitem [{\citenamefont {H\"{u}bner}\ \emph {et~al.}(2013)\citenamefont
  {H\"{u}bner}, \citenamefont {Lonnemann}, \citenamefont {Zell}, \citenamefont
  {Kuhn}, \citenamefont {Berski},\ and\ \citenamefont
  {Oestreich}}]{Hubner2013}%
  \BibitemOpen
  \bibfield  {author} {\bibinfo {author} {\bibfnamefont {J.}~\bibnamefont
  {H\"{u}bner}}, \bibinfo {author} {\bibfnamefont {J.~G.}\ \bibnamefont
  {Lonnemann}}, \bibinfo {author} {\bibfnamefont {P.}~\bibnamefont {Zell}},
  \bibinfo {author} {\bibfnamefont {H.}~\bibnamefont {Kuhn}}, \bibinfo {author}
  {\bibfnamefont {F.}~\bibnamefont {Berski}}, \ and\ \bibinfo {author}
  {\bibfnamefont {M.}~\bibnamefont {Oestreich}},\ }\href {\doibase
  10.1364/OE.21.005872} {\bibfield  {journal} {\bibinfo  {journal} {Opt. Exp.}\
  }\textbf {\bibinfo {volume} {21}},\ \bibinfo {pages} {5872} (\bibinfo {year}
  {2013})}\BibitemShut {NoStop}%
\bibitem [{\citenamefont {Berski}\ \emph {et~al.}(2013)\citenamefont {Berski},
  \citenamefont {Kuhn}, \citenamefont {Lonnemann}, \citenamefont {H\"{u}bner},\
  and\ \citenamefont {Oestreich}}]{Berski2013a}%
  \BibitemOpen
  \bibfield  {author} {\bibinfo {author} {\bibfnamefont {F.}~\bibnamefont
  {Berski}}, \bibinfo {author} {\bibfnamefont {H.}~\bibnamefont {Kuhn}},
  \bibinfo {author} {\bibfnamefont {J.~G.}\ \bibnamefont {Lonnemann}}, \bibinfo
  {author} {\bibfnamefont {J.}~\bibnamefont {H\"{u}bner}}, \ and\ \bibinfo
  {author} {\bibfnamefont {M.}~\bibnamefont {Oestreich}},\ }\href {\doibase
  10.1103/PhysRevLett.111.186602} {\bibfield  {journal} {\bibinfo  {journal}
  {Phys. Rev. Lett.}\ }\textbf {\bibinfo {volume} {111}},\ \bibinfo {pages}
  {186602} (\bibinfo {year} {2013})}\BibitemShut {NoStop}%
\bibitem [{\citenamefont {Torrey}(1956)}]{Torrey1956}%
  \BibitemOpen
  \bibfield  {author} {\bibinfo {author} {\bibfnamefont {H.}~\bibnamefont
  {Torrey}},\ }\href {\doibase 10.1103/PhysRev.104.563} {\bibfield  {journal}
  {\bibinfo  {journal} {Phys. Rev.}\ }\textbf {\bibinfo {volume} {104}},\
  \bibinfo {pages} {563} (\bibinfo {year} {1956})}\BibitemShut {NoStop}%
\bibitem [{\citenamefont {Qi}\ and\ \citenamefont {Zhang}(2003)}]{Qi2003}%
  \BibitemOpen
  \bibfield  {author} {\bibinfo {author} {\bibfnamefont {Y.}~\bibnamefont
  {Qi}}\ and\ \bibinfo {author} {\bibfnamefont {S.}~\bibnamefont {Zhang}},\
  }\href {\doibase 10.1103/PhysRevB.67.052407} {\bibfield  {journal} {\bibinfo
  {journal} {Phys. Rev. B}\ }\textbf {\bibinfo {volume} {67}},\ \bibinfo
  {pages} {052407} (\bibinfo {year} {2003})}\BibitemShut {NoStop}%
\bibitem [{Note1()}]{Note1}%
  \BibitemOpen
  \bibinfo {note} {\label {nt: supMat}See Appendices for a schematic of the
  optical path and equation derivations.}\BibitemShut {Stop}%
\bibitem [{\citenamefont {Dzhioev}\ \emph {et~al.}(2002)\citenamefont
  {Dzhioev}, \citenamefont {Kavokin}, \citenamefont {Korenev}, \citenamefont
  {Lazarev}, \citenamefont {Meltser}, \citenamefont {Stepanova}, \citenamefont
  {Zakharchenya}, \citenamefont {Gammon},\ and\ \citenamefont
  {Katzer}}]{Dzhioev2002}%
  \BibitemOpen
  \bibfield  {author} {\bibinfo {author} {\bibfnamefont {R.~I.}\ \bibnamefont
  {Dzhioev}}, \bibinfo {author} {\bibfnamefont {K.~V.}\ \bibnamefont
  {Kavokin}}, \bibinfo {author} {\bibfnamefont {V.~L.}\ \bibnamefont
  {Korenev}}, \bibinfo {author} {\bibfnamefont {M.~V.}\ \bibnamefont
  {Lazarev}}, \bibinfo {author} {\bibfnamefont {B.~Y.}\ \bibnamefont
  {Meltser}}, \bibinfo {author} {\bibfnamefont {M.~N.}\ \bibnamefont
  {Stepanova}}, \bibinfo {author} {\bibfnamefont {B.~P.}\ \bibnamefont
  {Zakharchenya}}, \bibinfo {author} {\bibfnamefont {D.}~\bibnamefont
  {Gammon}}, \ and\ \bibinfo {author} {\bibfnamefont {D.~S.}\ \bibnamefont
  {Katzer}},\ }\href {\doibase 10.1103/PhysRevB.66.245204} {\bibfield
  {journal} {\bibinfo  {journal} {Phys. Rev. B}\ }\textbf {\bibinfo {volume}
  {66}},\ \bibinfo {pages} {245204} (\bibinfo {year} {2002})}\BibitemShut
  {NoStop}%
\bibitem [{\citenamefont {Kikkawa}\ and\ \citenamefont
  {Awschalom}(1998)}]{Kikkawa1998}%
  \BibitemOpen
  \bibfield  {author} {\bibinfo {author} {\bibfnamefont {J.~M.}\ \bibnamefont
  {Kikkawa}}\ and\ \bibinfo {author} {\bibfnamefont {D.~D.}\ \bibnamefont
  {Awschalom}},\ }\href {\doibase 10.1103/PhysRevLett.80.4313} {\bibfield
  {journal} {\bibinfo  {journal} {Phys. Rev. Lett.}\ }\textbf {\bibinfo
  {volume} {80}},\ \bibinfo {pages} {4313} (\bibinfo {year}
  {1998})}\BibitemShut {NoStop}%
\bibitem [{\citenamefont {Kuhlen}\ \emph {et~al.}(2014)\citenamefont {Kuhlen},
  \citenamefont {Ledesch}, \citenamefont {de~Winter}, \citenamefont
  {Althammer}, \citenamefont {G\"{o}nnenwein}, \citenamefont {Opel},
  \citenamefont {Gross}, \citenamefont {Wassner}, \citenamefont {Brandt},\ and\
  \citenamefont {Beschoten}}]{Kuhlen2014}%
  \BibitemOpen
  \bibfield  {author} {\bibinfo {author} {\bibfnamefont {S.}~\bibnamefont
  {Kuhlen}}, \bibinfo {author} {\bibfnamefont {R.}~\bibnamefont {Ledesch}},
  \bibinfo {author} {\bibfnamefont {R.}~\bibnamefont {de~Winter}}, \bibinfo
  {author} {\bibfnamefont {M.}~\bibnamefont {Althammer}}, \bibinfo {author}
  {\bibfnamefont {S.~T.~B.}\ \bibnamefont {G\"{o}nnenwein}}, \bibinfo {author}
  {\bibfnamefont {M.}~\bibnamefont {Opel}}, \bibinfo {author} {\bibfnamefont
  {R.}~\bibnamefont {Gross}}, \bibinfo {author} {\bibfnamefont {T.~a.}\
  \bibnamefont {Wassner}}, \bibinfo {author} {\bibfnamefont {M.~S.}\
  \bibnamefont {Brandt}}, \ and\ \bibinfo {author} {\bibfnamefont
  {B.}~\bibnamefont {Beschoten}},\ }\href {\doibase 10.1002/pssb.201350201}
  {\bibfield  {journal} {\bibinfo  {journal} {Physica Status Solidi (B)}\
  }\textbf {\bibinfo {volume} {251}},\ \bibinfo {pages} {1861} (\bibinfo {year}
  {2014})}\BibitemShut {NoStop}%
\bibitem [{\citenamefont {Baumberg}\ \emph {et~al.}(1994)\citenamefont
  {Baumberg}, \citenamefont {Crooker}, \citenamefont {Awschalom}, \citenamefont
  {Samarth}, \citenamefont {Luo},\ and\ \citenamefont
  {Furdyna}}]{Baumberg1994}%
  \BibitemOpen
  \bibfield  {author} {\bibinfo {author} {\bibfnamefont {J.~J.}\ \bibnamefont
  {Baumberg}}, \bibinfo {author} {\bibfnamefont {S.~A.}\ \bibnamefont
  {Crooker}}, \bibinfo {author} {\bibfnamefont {D.~D.}\ \bibnamefont
  {Awschalom}}, \bibinfo {author} {\bibfnamefont {N.}~\bibnamefont {Samarth}},
  \bibinfo {author} {\bibfnamefont {H.}~\bibnamefont {Luo}}, \ and\ \bibinfo
  {author} {\bibfnamefont {J.~K.}\ \bibnamefont {Furdyna}},\ }\href {\doibase
  10.1103/PhysRevB.50.7689} {\bibfield  {journal} {\bibinfo  {journal} {Phys.
  Rev. B}\ }\textbf {\bibinfo {volume} {50}},\ \bibinfo {pages} {7689}
  (\bibinfo {year} {1994})}\BibitemShut {NoStop}%
\bibitem [{\citenamefont {Oestreich}\ and\ \citenamefont
  {R\"{u}hle}(1995)}]{Oestreich1995}%
  \BibitemOpen
  \bibfield  {author} {\bibinfo {author} {\bibfnamefont {M.}~\bibnamefont
  {Oestreich}}\ and\ \bibinfo {author} {\bibfnamefont {W.~W.}\ \bibnamefont
  {R\"{u}hle}},\ }\href {\doibase 10.1103/PhysRevLett.74.2315} {\bibfield
  {journal} {\bibinfo  {journal} {Phys. Rev. Lett.}\ }\textbf {\bibinfo
  {volume} {74}},\ \bibinfo {pages} {2315} (\bibinfo {year}
  {1995})}\BibitemShut {NoStop}%
\bibitem [{\citenamefont {H\"{u}bner}\ \emph {et~al.}(2009)\citenamefont
  {H\"{u}bner}, \citenamefont {D\"{o}hrmann}, \citenamefont {H\"{a}gele},\ and\
  \citenamefont {Oestreich}}]{Hubner2009}%
  \BibitemOpen
  \bibfield  {author} {\bibinfo {author} {\bibfnamefont {J.}~\bibnamefont
  {H\"{u}bner}}, \bibinfo {author} {\bibfnamefont {S.}~\bibnamefont
  {D\"{o}hrmann}}, \bibinfo {author} {\bibfnamefont {D.}~\bibnamefont
  {H\"{a}gele}}, \ and\ \bibinfo {author} {\bibfnamefont {M.}~\bibnamefont
  {Oestreich}},\ }\href {\doibase 10.1103/PhysRevB.79.193307} {\bibfield
  {journal} {\bibinfo  {journal} {Phys. Rev. B}\ }\textbf {\bibinfo {volume}
  {79}},\ \bibinfo {pages} {193307} (\bibinfo {year} {2009})}\BibitemShut
  {NoStop}%
\bibitem [{\citenamefont {Yu}\ and\ \citenamefont {Cardona}(2010)}]{Yu2010}%
  \BibitemOpen
  \bibfield  {author} {\bibinfo {author} {\bibfnamefont {P.~Y.}\ \bibnamefont
  {Yu}}\ and\ \bibinfo {author} {\bibfnamefont {M.}~\bibnamefont {Cardona}},\
  }\href {\doibase 10.1007/978-3-642-00710-1} {\emph {\bibinfo {title}
  {{Fundamentals of Semiconductors}}}},\ Graduate Texts in Physics\ (\bibinfo
  {publisher} {Springer Berlin Heidelberg},\ \bibinfo {address} {Berlin,
  Heidelberg},\ \bibinfo {year} {2010})\BibitemShut {NoStop}%
\end{thebibliography}%

\appendix

\newpage

\setcounter{figure}{0}
\renewcommand{\figurename}{FIG. S}
\begin{figure}
	\includegraphics[scale=0.5]{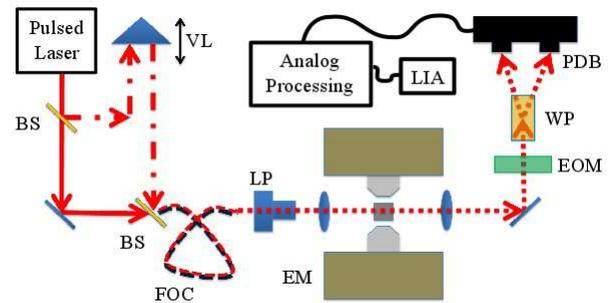}
	\caption{Optical path described in the text.  BS = beam splitter, EM = electromagnet, EOM = electro-optic modulator, FOC = fiber optic cable, LIA = lock-in amplifier, LP = linear polarizer, PDB = photodiode bridge, VL = variable length path, WP = Wollaston Prism.}
	\label{fig: SupFig1}
\end{figure}

\section{Derivation of $\left\langle \theta_i \theta_j \right\rangle$ for spin dependent behavior in n-GaAs}

We describe the measured spin physics of our sample using Eq \ref{eq: supEq1}:
\begin{equation}
\frac{\partial \mathbf{S}}{\partial t} - \mathbf{\Omega} \times \mathbf{S} + \frac{\mathbf{S}}{\tau_S} = \mathbf{S_0} \delta \left( t \right)
\label{eq: supEq1}
\end{equation}
where $\mathbf{S}$ is the time-dependent spin density, $\mathbf{S_0}$ is the initial spin density, $\mathbf{\Omega}$ is the Larmor precession frequency as defined in the Letter, $\tau_S$ is the spin dephasing time, and $\delta \left( t \right)$ is the Dirac-delta function.  The difference between the spin up and spin down components of the carrier density $n$ along the optical path is defined as $n_S \equiv n_\uparrow - n_\downarrow = \vert \mathbf{S_0} \vert$ with $n=n_\uparrow +n_\downarrow$.  Equation \ref{eq: supEq1} is an isotropic version of the Bloch-Torrey equations that excludes drift and diffusion and is applicable to our n-GaAs sample and experiment conditions.  We have incorporated  $\delta \left( t \right)$ so that we may solve for the Green's function.

\begin{figure}[h]
	\includegraphics[scale=0.7]{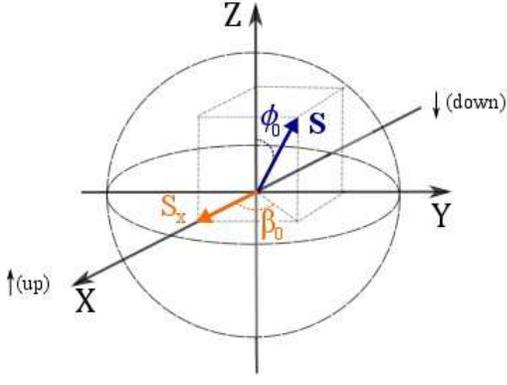}
	\caption{Sphere for the vectorial representation of the spin density vector.}
	\label{fig: SupFig2}
\end{figure}

Using a sphere (see Fig. S\ref{fig: SupFig2}), we define the spin projection along the optical path to be in the x-direction with an applied magnetic field in the z-direction yielding:
\begin{equation}
S_x = n_S H\left(t-t_0\right) e^{\left(t-t_0\right)/\tau_S} \sin \phi_0 \cos \left[\Omega \left(t - t_0 \right) \right]
\label{eq: supEq2}
\end{equation}

where $\phi_0$ and $\beta_0$ are the initial orientation of the spin polarization.  $H\left(t-t_0 \right)$ is the Heaviside step function with $t_0$ being the initial time of generation for the spin population.

We let $t-t_0 \rightarrow t_i$ and $S_x \rightarrow S_{xi}$ be a random time and associated x-projection of the spin polarization respectively.  We also ignore the Heaviside function by assuming repeated sampling of a single spin population that was polarized at a random instant in time.  We then average over all configurations of $n_S,\beta_0,$ and $\phi_0$.  As in Eq. \ref{eq: expectation-of-product}, we obtain the correlation from the expectation of the product $S_{xi} S_{xj}$
\begin{equation}
\left\langle S_{xi} S_{xj} \right\rangle = \left\langle S_{xi} \right\rangle\left\langle S_{xj} \right\rangle + Var\left( S_x \right) Corr \left( S_{xi}, S_{xj} \right)
\label{eq: supEq3}
\end{equation}
For a sample with zero net spin polarization, $\left\langle S_{xi} \right\rangle = \left\langle S_{xj} \right\rangle = 0$, and $\left\langle S_{xi} \right\rangle \left\langle S_{xj} \right\rangle = 0$.  We write the expectation value of the product of two ensemble spin states as:
\begin{align}
&\left\langle S_{xi} S_{xj} \right\rangle = \nonumber \\
& \quad \left\langle n_S^2 e^{\left(t_i + t_j\right)/\tau_S} \sin^2 \phi_0 \cos \left[\Omega t_i + \beta_0 \right] \cos \left[\Omega t_j + \beta_0 \right] \right\rangle
\label{eq: supEq4}
\end{align}
Only the relative time, not the absolute time, between states is of importance so we let $t_j \rightarrow t_i+\vert \Delta t \vert$.  We also let $t_i \rightarrow 0$ as the absolute time should not affect the value of the correlation between measurements, yielding:
\begin{equation}
\left\langle S_{xi} S_{xj} \right\rangle = \left\langle n_S^2 e^{\vert \Delta t \vert /\tau_S} \sin^2 \phi_0 \cos \beta_0 \cos \left[\Omega \vert \Delta t \vert + \beta_0 \right] \right\rangle
\label{eq: supEq5}
\end{equation}

We can relate the measured Faraday rotation to the ensemble spin polarization using $\theta = \sigma_F n_S d$.  We can also simplify our expression by evaluating the average of $\phi_0 \in [0,\pi)$ and $\beta_0 \in [0,2\pi)$ which results in the factor 1/4.  Our result is
\begin{equation}
\left\langle \theta_{i} \theta_{j} \right\rangle = \frac{1}{4} \left(\sigma_F d
\right)^2 \left\langle n_S^2 \right\rangle e^{\vert \Delta t \vert /\tau_S} \cos \left(\Omega \vert \Delta t \vert \right)
\label{eq: supEq6}
\end{equation}
where we recognize $Var(\theta_i)=\frac{1}{4} \left(\sigma_F d
\right)^2 \left\langle n_S^2 \right\rangle$ and $Corr(\theta_i,\theta_j)= e^{\vert \Delta t \vert /\tau_S} \cos \left(\Omega \vert \Delta t \vert \right)$.  

The last portion of our derivation involves the evaluation of $\left\langle n_S^2 \right\rangle$, the radial magnitude of the spin polarization.  Using the fact that spins are a two state system, we can evaluate $\left\langle n_S^2 \right\rangle$ as the variance of a binomial distribution.  Therefore $\left\langle n_S^2 \right\rangle = N/4V^2$ where $N$ is the total number of probed spins and $V=Ad$ is the probed volume equivalent to the probe cross-section multiplied by the thickness of the sample.  If we recognize that $n=N/Ad$, we arrive at our final expression for the variance:

\begin{equation}
Var\left(\theta_i\right) = \frac{\sigma_F}{16} \frac{n d}{A}
\label{eq: supeq7}
\end{equation}

\section{Discussion of weighted average for two pulse trains with variable delay}

\begin{figure}[h]
	\includegraphics[scale=0.7]{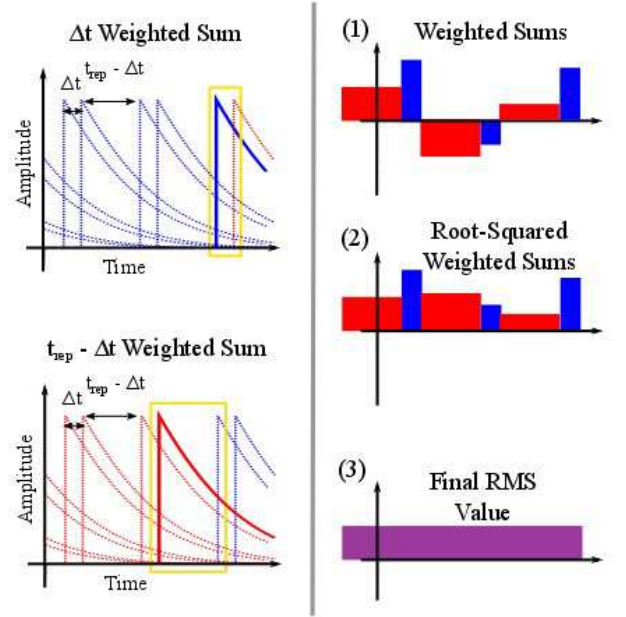}
	\caption{Schematic representation of electrical processing.}
	\label{fig: SupFig3}
\end{figure}

We begin with a schematic representation of the electrical processing.  On the left in Fig. S\ref{fig: SupFig3}, the yellow boxes highlight the two possible scenarios:  summing over all previous pulses while averaging the decay over an interval of $\Delta t$ (top, series in blue); summing over all previous pulses while averaging over the decay over an interval of $t_{rep}-\Delta t$ (bottom, series in red).  We treat the electrical pulses generated at the photodiode bridge as having an instantaneous rise time (3 ps Ti:Saph laser pulsewidth) and a finite fall time due to the limited response of the electronics ($t_{RC} \sim$ 29 ns).
	
Since our electronics are in series, the processing is first-in-first-out which we schematically follow on the right side of Fig. S\ref{fig: SupFig3}.  First, a weighted sum over either the blue series or the red series is performed.  Second, the value of the weighted sum is squared and then square-rooted.  Third, an average over many weighted sums is performed leading to a steady state value which we extract using lock-in detection.

We can write down the mathematics in a straightforward fashion.  For the squared pulse response weighted sum, we write

\begin{equation}
f_{\alpha,\beta} = \left[\frac{1}{\beta} \int_{0}^{\beta} dt \, e^{-t/t_{RC}}\right]^2 \frac{\sum_{i,j}^{\infty} e^{-\left(t_i^{\alpha}+t_j^{\alpha}\right)/t_{RC}} \left\langle \theta_i^{\alpha} \theta_j^{\alpha}\right\rangle }{\sum_{i,j}^{\infty} e^{-\left(t_i^{\alpha}+t_j^{\alpha}\right)/t_{RC}}}
\label{eq: supEq8}
\end{equation}

Equation \ref{eq: supEq8} is a modified version of Eq. 3 in the paper where the time-points $t_n^{\alpha}$ are not evenly spaced.  For $f_{\alpha,\beta}, \beta$ is the period for calculating a given weighted sum while $\alpha$ is the first non-zero time-point in the series of sampled times $t_i^{\alpha}\in \lbrace0,\alpha,t_{rep},t_{rep+\alpha},…\rbrace$.  For our experiment, $\beta=\Delta t$ while $\alpha= t_{rep}-\Delta t$ with $f_{\alpha,\beta}$ corresponding to the blue series and $f_{\beta,\alpha}$ (the permutation of $\alpha$ and $\beta$ values) to the red series shown in Fig. S\ref{fig: SupFig3}.

As shown on the bottom right hand side of Fig. S\ref{fig: SupFig3}, we only measure a single value as our electronics perform a square-root, followed by a weighted average, of $f_{\alpha,\beta}$ and $f_{\beta,\alpha}$.  We write this mathematically as:
\begin{equation}
\left\langle \theta_{WA}^2 \right\rangle = \frac{\beta}{t_{rep}}\sqrt{f_{\alpha,\beta}} + \frac{\alpha}{t_{rep}}\sqrt{f_{\beta,\alpha}}
\label{eq: supEq9}
\end{equation}
where we make use of the fact that $\alpha+\beta=t_{rep}$.  The weighting factor for each series is determined by the relative interval lengths of the respective weighted sums.

\section{Comparison to Model Used by Ref. 21}

\begin{figure}
	\includegraphics[scale=1]{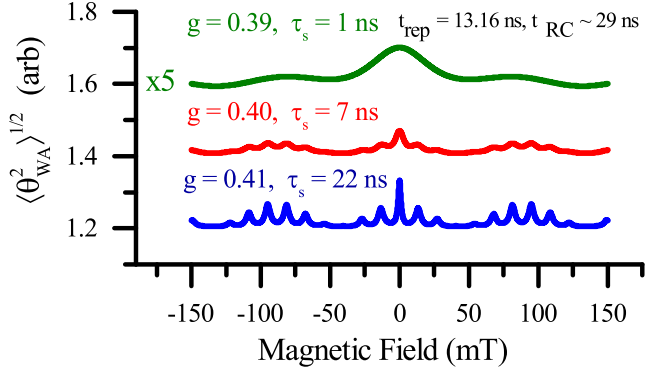}
	\caption{Plot of combined Eqs. \ref{eq: supEq8}, \ref{eq: supEq9}, and \ref{eq: supEq10} for the same parameters as in Fig. \ref{fig: Fig2} in our paper.  Equation \ref{eq: supEq10} comes from taking the antiderivative of Eq. 1 in Ref. 21
	.  As shown in our paper, we do not observe a pronounced peak at 0-field in our data set.}
	\label{fig: SupFig4}
\end{figure}

If we take the antiderivative of Eq. 1 in Ref. 21\nocite{Berski2013a}, we arrive at:
\begin{eqnarray}
&\left\langle \theta_i \theta_j \right\rangle \sim
\frac{1}{1 + \tau_s^2 \Omega^2} e^{-\vert \Delta t\vert / \tau_S} \qquad \qquad \qquad \qquad \qquad \qquad \qquad \quad \nonumber \\ 
& \; \times \left(\left( 2 + \tau_S^2 \Omega^2 \right) \cos\left[ \Omega \vert \Delta t \vert \right] - \tau_S \Omega \sin\left[ \Omega \vert \Delta t \vert \right] \right)
\label{eq: supEq10}
\end{eqnarray}
This would predict a Lorentzian type scaling near zero magnetic field for a Resonant Spin Noise scan.  If we use this model for Eqs. \ref{eq: supEq8} and \ref{eq: supEq9}, we would expect the signal plotted in Fig. S\ref{fig: SupFig4}.

Reference 21\nocite{Berski2013a} does not include a full derivation of their model.  However, Eq. \ref{eq: supEq10} can be derived by calculating the integral
\begin{eqnarray}
\int_0^{\infty} dt \, f\left( t \right) f\left( t- \vert \Delta t\vert \right) \label{eq: supEq11a} \\
f\left(t\right) \equiv e^{-t/\tau_S} \cos\left[\Omega t\right] \label{eq: supEq11b}
\end{eqnarray}
Equation \ref{eq: supEq11a} is the autocorrelation of $f\left(t\right)$, but the result (Eq. \ref{eq: supEq10}) does not appear to accurately describe the measured spin noise behavior.  In our model (Eq. \ref{eq: supEq6}), the spin ensemble correlation function is Eq. \ref{eq: supEq11b} and not its autocorrelation.

\end{document}